%% file: main.tex
\documentclass[sigconf, 10pt]{acmart}
\usepackage[english]{babel}
\usepackage{blindtext}

\usepackage{soul}
\usepackage{pifont}
\usepackage[medium,compact]{titlesec}
\usepackage{xurl}

\input{new-commands}

\newcommand{\roce}{RoCEv2\xspace}
\newcommand{\sys}{STrack\xspace}
\newcommand{\mpath}{multipath\xspace}
\newcommand{\subsec}{Section\xspace}
\newcommand{\algo}{Algo\xspace}
\newcommand{\tbl}{Table\xspace}
\newcommand{\fig}{Figure\xspace}
\newcommand{\evenspray}{oblivious packet spraying\xspace}
\newcommand{\adptivespray}{adaptive packet spraying\xspace}
\newcommand{\dbt}{DoubleBinary Tree\xspace}
\newcommand{\qpsperconn}{QPS\_PER\_CONN\xspace}

\titlespacing*{\section}{1pt}{3.5pt}{2pt}
\titlespacing*{\subsection}{1pt}{3pt}{1.5pt}
\titlespacing*{\subsubsection}{1pt}{3pt}{1.5pt}

\usepackage{tikz}

\newcommand{\yanfang}[1]{\textbf{\color{red}yle: #1}}

\renewcommand\footnotetextcopyrightpermission[1]{} 

\setcopyright{none}
\acmPrice{}

\begin{document}

\title{\sys: A Reliable Multipath Transport for AI/ML Clusters}

\author{Yanfang Le$^1$,
Rong Pan$^1$,
Peter Newman$^1$,
Jeremias Blendin$^1$,\\
Abdul Kabbani$^2$,
Vipin Jain$^1$,
Raghava Sivaramu$^1$, 
Francis Matus$^1$
}
\affiliation{$^1$AMD\;$^2$Microsoft}

\renewcommand{\shortauthors}{Y. Le et al.}

\begin{abstract}
Emerging artificial intelligence (AI) and machine learning (ML) workloads present new challenges of managing the collective communication used
in distributed training across hundreds or even thousands of GPUs.
This paper presents \sys, a novel hardware-offloaded reliable transport protocol aimed at improving the performance of AI /ML workloads by rethinking key aspects of the transport layer. \sys optimizes congestion control and load balancing in tandem: it incorporates an adaptive load balancing algorithm leveraging ECN, while adopts RTT as multi-bit congestion indicators for precise congestion window adjustment. Additionally, STrack facilitates out-of-order delivery, selective retransmission, and swift loss recovery in hardware for multipath environment.
The extensive simulation comparing \sys and \roce demonstrates that \sys outperforms \roce by up to $6$X with synthetic workloads and by $27.4$\% with collective workloads, even with the optimized \roce system setup.
\end{abstract}

\maketitle

\input{sections/1intro}
\input{sections/2motivation}

\input{sections/3design}  
\input{sections/5evaluation}
\input{sections/6related}
\input{sections/7con}
\input{sections/acknowledgement}

\bibliographystyle{plain}
\bibliography{refer}

\end{document}

%% file: new-commands.tex
\usepackage{url}
\usepackage{graphicx}
\usepackage{latexsym}
\usepackage{amsmath}
\usepackage{bm}
 
\usepackage{amssymb}
\usepackage{amsfonts}
\usepackage{comment}
\usepackage{array}
\usepackage{xspace}
\usepackage{psfrag}
\usepackage[tight,small]{subfigure}
\usepackage{color}
\usepackage{algorithm}
\usepackage{algorithmicx}
\usepackage{algpseudocode}
\usepackage{enumitem}
\usepackage{alltt}
\usepackage{multirow}
\usepackage{array}
\usepackage{wasysym}
\usepackage{rotating}
\usepackage{datetime}
\usepackage{amsthm}
\usepackage{epsfig}
\usepackage{endnotes}
\usepackage{epstopdf}
\usepackage{booktabs}

%% file: sections/1intro.tex
\section{Introduction}

The widely adopted transport technology in the AI/ML cluster today is Infiniband. However, to achieve even larger scale with lower cost, leading CSPs start to adopt Ethernet based technology, such as RDMA over Converged Ethernet, RoCEv2~\cite{rocev2}. In traditional cloud environment, RoCEv2 is proven to provide ultra-low latency and high throughput with little CPU overhead. However, the standard RoCEv2 has many limitations to be adopted widely for AI/ML clusters. 

First, AI/ML training networks demand much higher utilization of their network links, around 70\%-80\% while cloud networks usually only carry a load of 30\%-50\%. Due to ECMP hash collisions, \roce, as a single-path transport, fails to evenly distribute traffic load across highly parallel paths in the AI/ML network, and  can't provide robustness in the face of link failures. As a result, hot spots are congested while other links run empty. The overall performance suffers. Hence, RoCEv2 can't satisfy the high link utilization requirement of AI/ML workloads. 

Secondly, as AI/ML cluster networks grow to connect tens of thousands or hundreds of thousands nodes, transport's error resiliency capability is a must in order to achieve high efficiency. Multipathing would make error recovery even more challenging as the transport needs to determine whether a missing packet is a real packet loss or simply a delayed packet due to multipathing. Although RoCEv2's dependency on lossless Ethernet can help prevent congestion packet drops, bit errors or packet delays happen often in a vast AI/ML network. RDMA's go-back-N error recovery mechanism would result in significant performance loss. 

Thirdly, highly paralleled paths in AI/ML clusters demand a congestion control algorithm that works jointly with a load balancing scheme. It is a dedicated act of quickly switching paths when there are under-utilized links versus slowing down in face of heavy congestion across paths. This calls for a new approach to congestion control with the consideration about the effect of adaptive load balancing. DCQCN, the de facto congestion control algorithm for RoCEv2, fails to satisfy such a stringent requirement. In addition to being a single path scheme, DCQCN is a rate-based algorithm that interprets lacking of congestion notifications as good network conditions, which often exacerbates congestion level. This would lead to extremely unbalanced network links, not desirable for AI/ML workloads. 

Last but not the least, to achieve the low latency requirement of the AI/ML workloads, it would be ideal to continue hardware offloading the transport layer. To support multipathing, we need to track the congestion
states in hardware on each path in order to enable
congestion-aware load distribution. However, these
states grow linearly with the number of sending paths.
This could cause a considerable memory overhead even
when a modest number of paths are used. In addition, multipathing can cause the out-of-order delivery at NICs, even if we directly DMA data packets to hosts, the memory footprint to track the packet arrival and control information are still needed at the NIC's cache. To minimize the states required, the congestion control and load balancing need to work together and minimize the out-of-order delivery. DCQCN's inability to effectively balance traffic load would require large re-ordering states at receiver NICs.

This paper presents \sys, the first reliable multipath transport that addresses all aforementioned
challenges. \sys employs a novel
mechanism that adaptively spray packets to multiple
paths without keeping complicated per path state. In addition, we design a window-based congestion control that yields to path selection choice first before cutting down the window in face of pending congestion. We also assume lossy Ethernet as the link layer technology, and design a reliable error recovery mechanism that is based on out-of-order packet counts at receiver NICs to ensure fast packet recovery with minimal spurious retransmission.

\sys uses egress-marked ECN as a congestion signal for the adaptive packet spray algorithm instead of oblivious packet spray that distributes packets across multiple paths evenly. Note that the paths here are not the physical paths, but different entropy values, e.g. different UDP source ports, that a sender NIC's uses to spray traffic. With different entropies in the packet header, ECMP functions at switches would hash them onto different paths. However, oblivious packet spray still experiences hash collisions, and the situation gets worse over time as congested paths accumulate packets as messages continue. In addition, link failures cause full bisectional network becomes asymmetric, which makes oblivious packet spray unbalanced.  \sys keeps a simple bitmap for the entropies that have experienced ECN marks as the congestion state. When an ACK comes back without an ECN mark, that entropy is used to clock out the next new packet. If an ACK comes back with an ECN marked, it is marked in the corresponding position in the bitmap. Next non-marked entropy in a round robin manner is used to clock out new packet. Note that S-track uses only a minimal state, bitmap, to keep congestion information. The bitmap is reset after one or two round trip times.

\sys adopts a sender-only congestion control to handle both fabric congestion and last hop incast. Although a receiver-based scheme can handle a large degree of incast, it is well-known that the AI/ML communication collectives are designed to avoid incast at the application layer. The transport layer incasts do occur but to a much less degree, only up to tens of flows, where sender-based algorithm can easily deal with. In addition, receiver-based congestion control requires sender-based design to combat fabric congestion, and adds more memory footprint to the receiver, which conflicts with our goal of reducing hardware offload complexity.

\sys also addresses RoCEv2's inefficiency  by using a window-based algorithm that handles the congestion across the paths and works jointly with the adaptive load balancing scheme.
As the congestion shows up, ECN marks packets when they are at the egress, exiting the congested queue. This gives us an early indication of pending congestion. ECN-marked entropies are avoided by the sender. Unlike \cite{mprdma}, ECN-marked packets don't necessarily cause \sys to cut down its window. \sys's congestion control uses an average RTT, from the instantaneous RTT samples across different paths to decide whether we should cut the congestion window. Only when the average RTT is above a threshold (i.e., target queuing delay), \sys cuts the congestion window. The rational is that when only a few paths congested, we should take advantage of other paths without slowing down; and would cut rate when many paths are congested such as in an incast case. \sys also uses a novel method to quickly converge a flow's window under heavy congestion by utilizing total acknowledged bytes back at the sender. The details of our congestion control design can be found in Section~\ref{subsec:cc}.

At the receiver, \sys utilizes an coaleasing ACK approach to reduce the packets processing rate pressure on the NIC. It maintains a bitmap to track out-of-order packet arrival and selectively pick a bitmap segment (due to ACK size limit) to inform the sender to about the most-up-to-date view of the packet arrival. 
It also devises a novel silence packet loss detection mechanism across three different methods: (1)
out-of-order packets counters; (2) probing-based approach;
(3) timeout. 
Once packet loss is detected, the retransmited packets
are sent if the congestion window is allowed. 

We evaluate \sys with extensive simulations across a broad set of workloads, including permutations, incast traffic and collective workloads with different network scenarios, e.g., different link speed, link down, oversubscription networks, full bi-sectional networks. 
Our experiments results show that \sys outperforms \roce up to 
$6$X with synthetic workloads and by $27.4$\% with collective workloads, even with the optimized \roce system setup.

%% file: sections/2motivation.tex
\section{Background and Motivation}

\begin{figure*}[!htb]
    \centering
    \includegraphics[width=0.50\linewidth]{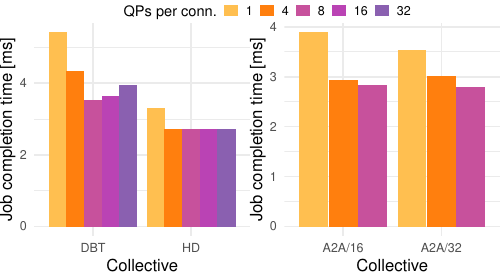}\hfill
    \includegraphics[width=0.45\linewidth]{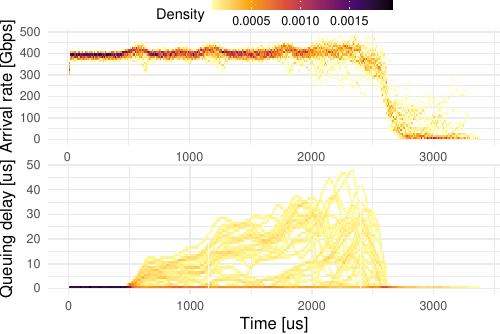}\hfill
    \begin{minipage}{0.24\textwidth}
        \centering
 		\caption{\small AllReduce collective finishing time.}
        \label{fig:ccl}
        \vfill
    \end{minipage}\hfill
    \begin{minipage}{0.24\textwidth}
        \centering
        \caption{\small AlltoAll collective finishing time. }
        \label{fig:a2a-motiv}
    \end{minipage}\hfill
    \begin{minipage}{0.40\textwidth}
        \centering
    \caption{\small \roce last-hop link queuing delay and arrival rate of A2A experiment.}
    \label{fig:dcqcnqdelay}
    \end{minipage} \hfill
\end{figure*}

\subsection{AI/ML Workloads Challenges over RoCEv2}
Collectives play a critical role in distributed AI/ML workloads by enabling efficient communication and synchronization among multiple computational nodes. They help in scaling the training processes, optimizing resource utilization, and reducing communication overhead, which are essential for handling the large-scale data and computational demands of modern AI/ML applications.

Collectives algorithms are known to be designed to avoid heavy incast, e.g., Ring AllReduce, Havingdoubling, \dbt ~\cite{nccl, rccl}, rely either one-to-one or two-to-one traffic patterns. Prior efforts~\cite{eflops, mprdma-aiml} have discovered the importance of the multi-path load balancing for AllReduce collectives. 
To illustrate the network load balancing impact on the collective communication, we run Allreduce collectives with \dbt (\textbf{DBT} in \fig \ref{fig:ccl}) and AlltoAll (\textbf{A2A} in \fig \ref{fig:ccl}) with different parallelization, i.e., $8$, $16$, and $32$. The network topology is a $2$-tier standard fat-tree topology with $32$ NICs to simulate $32$GPUs. The link speed is $400$Gbps and all the link speeds are the same in the network topology. The collective size is $128MB$.
Note that we run one collective at a time, i.e., one collective takes the whole cluster in these experiments. 

The collective completion time for Allreduce and AlltoAll operations, as shown in \fig \ref{fig:ccl}, indicates that using more entropies with \text{\roce, such as 4 \qpsperconn} (four QPs between source and destination), yields better performance than using single QP (\text{\roce, 1 \qpsperconn}). This suggests that ECMP hash collisions underutilize the paralleled path capacities of a network, highlighting the need for a more effective network load balancing scheme for AI/ML workloads.

Furthermore, different source of jitters in the system, e.g., stragglers, NIC scheduler or the GPU thread scheduler, can break the order of the collective algorithms, e.g., AlltoAll, and can cause the moderate incast. 
To demonstrate this point, \fig \ref{fig:dcqcnqdelay} monitors the queuing delay at all the last-hop switch queues (total of $32$ queues) and traffic arriving rate to each last-hop switch queue as time goes for the AlltoAll experiment with parallel degree of $32$ (A2A/32 in \fig \ref{fig:ccl}) for the \text{\roce, $1$ \qpsperconn} setup. 
Data shows that arrival rates at the last-hop queues exceed $400$ Gbps on multiple ocasions. For example, around $2$ ms, the arrival rate peaks at $489.256$Gbps at $2357$ $\mu$s. This high arrival rate, sustained for approximately $500\mu$s, results in switch queuing delays reaching up to $46\mu$s, equivalent to a queue depth of $2.3$MB. These significant queuing delays underscore the necessity for a more effective congestion control scheme tailored for AI/ML workloads.

In addition to the issues of single path load balancing and rate-based congestion control, \roce also requires network to be lossless due to its inefficient loss recovery mechanism, Go-Back-N. Lossless network usually is achieved by enabling priority flow control (PFC).
Prior works have been discussed that PFC storms~\cite{flor}, PFC caused network deadlock~\cite{rdmaatscale} and head-of-line blocking~\cite{irn, dcn}. In addition, given the vast scale of AI/ML clusters, link errors happen all the time. Adopting go-back-N to recover from link bit errors would be very inefficient. 
To eliminate the lossless network requirement, a selective retransmission mechanism is desired.

\subsection{Hardware Offloaded Transport}
\roce has been proven to achieve high-performance and low-latency by offloading the entire network stack to the NIC hardware. It directly transfers data to and from application buffer (e.g., GPU HBM), which bypasses time consuming processing in the CPU and saves the memory copy between host and GPU memory. 
To get the same benefits, it is desirable for a new transport for AI/ML workloads is hardware-offloaded as well. 

Fortunately, as AI/ML workloads do not have a strict requirement for in-order delivery. Similar as SRD~\cite{srd}, \sys provides reliable out-of-order delivery, without the need for a re-order buffer on the NIC.  
Nonetheless, the new transport must maintain states to support path congestion information for adaptive multipathing, out-of-order packet arrival information for selective acknowledgements and packet retransmissionson top of a new window-based congestion control algorithm. 
Accommodating this per-connection state in an efficient manner to allow a huge number of connections to be maintained in an hardware implementation is a key challenge.

\subsubsection{Multipath.}
Over the past decade, significant efforts has been made in improving load balancing in data center networks. 
There are generally two approaches: 1) one is to break down a single flow into multiple subflows, e.g. 4 or 8, and maintains a separate congestion window for each of the subflows, i.e., MPTCP~\cite{mptcp}; 2) the other approach is to spray packets across a large numer of paths, e.g. 128 or 256, and maintaining one congestion window across all paths \cite{srd}. The advantages of the subflow approach is that it keeps in-order packet delivery within each subflow. However, due to the per-path congestion state required, this approach limits the number of subflows that can be efficiently maintained in hardware in parallel.

If packets are sprayed over a large number of entropies (paths), the workload can be distributed more evenly. However, even with larger number of entropies, \evenspray can still perform poorly if the network is asymmetric and the hash collision can not be eliminated completely. Adaptive packet spray takes into account paths' congestion. The challenge is how to keep the path congestion state simple and efficient. If one congestion window governing traffic sent over all paralleled paths, the additional challenge  is how multipathing co-ordinates with the congestion control algorithm as various paths could experience different degrees of congestion.

\subsubsection{Congestion Management.}
Congestion management for AI/ML clusters is unique: the highly paralleled paths dictate a fine-grained packet spraying approach to achieve high network utilization. It also means that the traditional congestion control algorithms, which apply window control to a single path, can't be adopted here as each path often carries a single packet of a flow during an RTT. Hence, we need to design a mechanism that manages the overall traffic across many different paths or entropies and allows the sender to switch between different paths in order to fully utilize the available capacities across the entire network. 

Ideally, the first course of action upon a moderate congestion on a few paths should be shifting some traffic from the over-loaded path to the under-loaded ones. Congestion control should not step in and start cutting the window as reducing the window would limit the number of packets that can be sent to other under-utilized paths.  Instead, it diverts some traffic from the over-loaded paths to the under-loaded ones. Congestion control would be necessary during fabric congestion because of over-subscription, network asymmetry, or incast event when multiple senders send to a single receiver.

\begin{figure}[!t]
        \begin{minipage}{\columnwidth}
        \centering
        \includegraphics{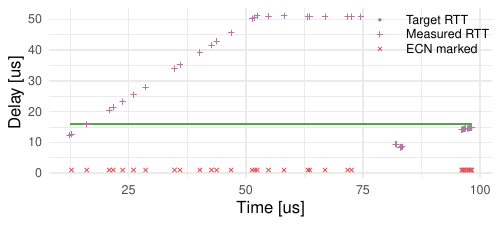}
 		\caption{Congestion Signals During an Incast.}
           \label{fig:cc-signal}
    \end{minipage}\hfill
\end{figure}
The design challenge is deciding when to switch paths and when to cut the window. We believe that the earliest congestion signal, indicating onset of congestion, should be used to change path. Without the assumption of network telemetry measures, we make the following key observation: modern switches adopt ECN marking at egress, when a packet exits a queue \cite{dcqcn}. This particular packet may not have experienced queueing delay, but its ECN mark indicates the queue behind it is building up. This would give us the earliest congestion signal, much faster than RTT or even the change of RTT. Figure \ref{fig:cc-signal} shows a simulation of a transient congestion event, when $32$ different sources send traffic to one receiver simultaneously. We measure ECN-marking as well as RTT as time progresses at a sender. As shown in Figure \ref{fig:cc-signal}, the first received ACK packet for this sender is already ECN-marked while the measured RTT is at base RTT, indicating this particular packet has not experienced any congestion but the queue behind it has built up. The measured RTT or any noticable RTT change does not happen until around 16us, much later after the ECN marked ACK packet has arrived.

With the above observation, we believe that ECN is the most timely signal for adaptive packet spraying. Only when a majority of paths experienced congestion, we would cut the window. As window adjustment is an quantitative decision, that's why we take RTT, a multi-bit signal, as the congestion signal for congestion control.

\begin{figure*}[!t]
    \begin{minipage}{0.25\textwidth}
        \centering
           \includegraphics[width=\linewidth]{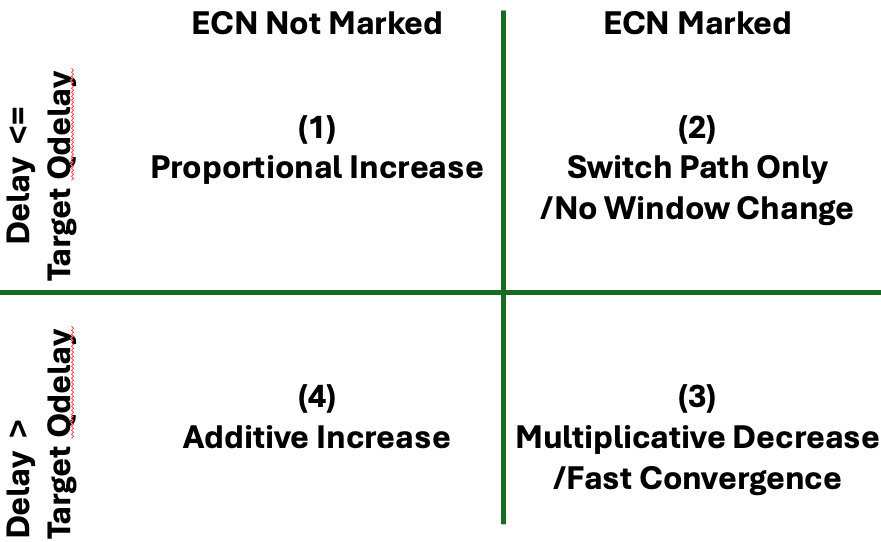}
 		\caption{\small Congestion Management Quadrant.}
           \label{fig:quad}
    \end{minipage}
    \begin{minipage}{0.24\textwidth}
        \centering
           \includegraphics[width=\linewidth]{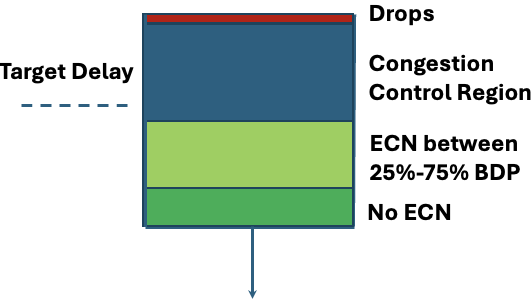}
 		\caption{\small Congestion Management Regions of Actions.}
           \label{fig:ecnth}
    \end{minipage}
        \hspace{1mm}
        \begin{minipage}{0.45\textwidth}
        \centering
    \includegraphics[width=\linewidth]{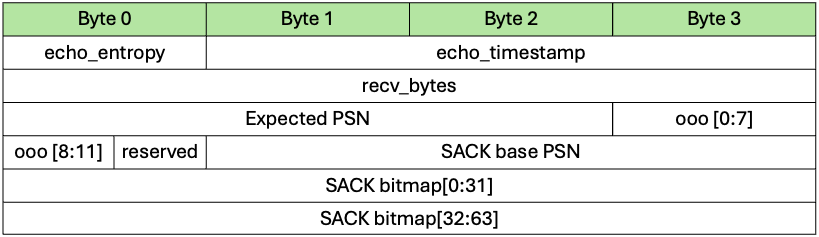}
 		\caption{\small Additional fields in ACK packet.}
           \label{fig:packet-format}
    \end{minipage}
\end{figure*}
\subsubsection{Reliability.}
Packets losses due to bit errors, buffer overflow, or device failures are common in data center networks. Transport layers uses packet retransmission to guarantee reliable packet delivery. A major departure from RoCEv2 in our approach is the assumption of lossy networks. In lossy networks, RoCEv2's Go-Back-N loss recovery mechanism becomes very ineffient. A mechanism that performs selective retransmission, i.e., only dropped packets are retransmited, is required. Selective retransmission requires timely ACK packets in order to track the packet receiving state. Due to multipathing, packet arrivals are often out of order. It is a challenge to detect packet losses quickly while minimizing spurious retransmissions. 

Ideally, acknowledging every arrival packet can provide the timely reliability information and does not require tracking states at a receiver. 
However, due to the NIC packet processing performance pressure as the link speed increases~\cite{srnic}, it is more efficent to coalease ACKs, i.e., the receivers generate ACKs after receiving a certain number of packets or due to certain special events. 

These coalesced ACKs relay packet arrival information back to the sender, referred to as \textbf{SACK}. Similar to previous works~\cite{irn, mprdma}, a bitmap to monitor the packet arrival status is often adopted. However, a single SACK packet cannot transmit the entire bitmap to the sender due to packet header size limitations. Consequently, selectively choosing which bitmap segment to send back to the sender to provide timely packet arrival information is one of the challenges that we must address.

Packet loss detection poses another issue in a multipath context, as the sender cannot determine whether a missing packet is simply delayed in the network or has been dropped. Timeout is a common method for detecting lost packets. However, timeout periods are often set high enough to account for queuing delays and unpredictable server/NIC processing time; otherwise, senders may experience spurious retransmissions. On the other hand, high timeout values can lead to undesirably long flow completion times. How to quickly and accurately detect silent packet losses is a major obstacle that needs to be solved.

%% file: sections/3design.tex

\section{\sys Design}

This section specifies three components of \sys's congestion management for reliable packet delivery: 1) adaptive packet spray in \subsec~\ref{subsec:app}; 2) sender-based congestion control for fabric and moderate scale incast congestion (up to $100$:$1$)  in \subsec~\ref{subsec:cc};  and 3) an selective retransmission algorithm in \subsec~\ref{subsec:lr} that takes into account out-of-order packets due to  \mpath. 
Our goal is to keep the design as simple as possible and does \textbf{not require} any advanced switch capabilities, such as Packet Trimming, Back-to-sender, In-network Telemetry; yet can work with them.

The congestion management in \sys handles congestion-aware multipathing and window-based congestion control jointly. Without the assumption of advanced switch features, \sys has two basic congestion signals to work with: a) egress-marked ECN; and b) RTT; Timely egress-marked ECN information is used to guide which paths shall be used or be avoided. The multi-bit congestion measures, RTT, is used to precisely control the congestion window. Based on these two signals, there are four scenarios that \sys handles as shown in Figure \ref{fig:quad}: 1) the acked packet is not ECN marked and its RTT is below the target RTT; 2) the acked packet is ECN marked but its RTT is still below the target RTT; 3) ECN is marked and its RTT is also above the target RTT; 4) ECN is not marked but its RTT is above the target RTT.

In Scenario \#1, the network is not congested, which is the most common network state. Hence, \sys would not interfere as long as the congestion window is at the maximum value, roughly the BDP. If the window is not at the max due to prior congestion episode, \sys increases the congestion window proportionally based on how far the RTT is from the target RTT. If the network is slightly congested, reflected by an marked ECN but with low RTT as in Scenario \#2, \sys chooses to switch path while keeping the congestion window intact to allow packets flow to other paths. When RTT becomes large indicating network wide congestion or incast scenarios, \sys cuts window multiplicatively in \#3. When RTT is much larger than the target, \sys uses additional congestion signal, achieved Bandwidth Delay Product (BDP), total acknowledged bytes in one base RTT, to quickly reaches the operating point in heavy incast scenarios. This allows \sys to converge much faster than the conventional AIMD approach. When congestion goes away, a packet that has experienced long queueing delay will not be ECN marked as no queue built up behind it. In this Scenario \#4, \sys actually starts increasing the window additively to avoid starving the link. 
The ECN marking threshold and the target RTT determine when \sys transitions from Scenario \#2 to Scenario \#3. 

The design has a clear separation between the congestion control and multipathing, i.e., the ECN marking threshold is between $25$\% and $75$\% of Bandwidth Delay Product (BDP) and the target queuing delay is one network RTT, as shown in Figure \ref{fig:ecnth}. Initially, we only switch paths based probabilistic ECN marking. As the network congestion increases, the majority of packets are ECN marked and their RTT increases which would trigger congestion window cut. The adaptive packet spray is still on-going even if with a window cut. Note that Figure \ref{fig:ecnth} reflects the symbolic queueing view from congestion control's angle while multiple paths' queue are involved in a real network.

\sys assumes that switches drop packets silently when no buffer space is available, i.e., trimming~\cite{ndp, cp} or Back-to-sender\cite{sfc, bolt} is not present. 
Furthermore, \sys's reliability mechanisms must also align with coalescing ACKs~\cite{smith2010tcp, johnson2018tcp} to accommodate the reality that packet processing rates grow at a slower pace than link speeds.
\sys provides a packet loss detection mechanism (\subsec ~\ref{subsec:lr}): (1) out-of-order packets counters; (2) probing-based approach; (3) timeout. Once loss is detected, the retransmited packets are sent if the congestion window is allowed.   

\begin{algorithm}[!t]
{\small
    \caption{\sys Overview}
    \label{algo:overview}
    \begin{algorithmic}[1]
    \vspace{1mm}
    \Procedure{\textbf{on\_sending\_packet}}{packet}
    \State packet.path\_id = choose\_path()
    \EndProcedure
    \Procedure{\textbf{on\_receiving\_ack}}{}
    \State ecn = ack.ecn 
    \State path\_id = ack.path\_id
    \State \textbf{update\_ecn\_bitmap(ecn, path\_id)}
    \If{measured\_rtt < base\_rtt}
    \State base\_rtt = measured\_delay
    \EndIf
    \State measured\_Qdelay = measured\_rtt - base\_rtt
    \State probe\_timer\_ts = now + 3* net\_base\_rtt
    \If{ack\_for\_probe $and$ delay $<$ $2*$ net\_base\_rtt $and$ achievedBDP == 0}
    \State \textbf{retransmit\_packets()}
    \EndIf
    \State achievedBDP = \textbf{update\_achievedBDP}(ack, now)
    \State \textbf{cwnd = adjust\_cwnd}(ecn, measured\_Qdelay, achievedBDP)
    \State \textbf{loss\_detection(ack)}
    \EndProcedure
    \end{algorithmic}
    }
        \end{algorithm}

\algo~\ref{algo:overview} shows the overall-design of \sys. We assume an ACK packet echoes back the ecn mark, path\_id and timestamp to the sender. Upon receiving an ACK, \sys checks: 1) whether an ECN is marked or not; 2) measured queuing delay (Line \#8); 3) Received ACKed bytes (Line \#13), which are our congestion signals. \sys updates the ECN bitmap with the echoed path\_id and ecn signal (Line \#4), which are used by multipath load balancing algorithm (in \algo ~\ref{algo:app}). The main congestion control algorithm \textbf{adjust\_cwnd} is specified in \algo ~\ref{algo:cc}, which we describe separately later.
The packet loss detection algorithm is described in \subsec~\ref{subsec:lr}.
Upon sending out a packet, we choose the right path id (\algo ~\ref{algo:app}), which is covered later.

\subsection{Adaptive Load Balancing}
\label{subsec:app}
\algo ~\ref{algo:app} explains \sys adaptive load balancing algorithm. It initializes a bitmap with all the paths available, i.e., set bit as $0$ and set the $next\_path\_id$ as invalid, i.e., a negative value.

Upon receiving an ACK packet, \sys remembers the entropy value that is returned ecn-free in next\_path\_id and label the path\_id as good path in the bitmap (Line \#7-8 in \algo \ref{algo:app}). 
When sending data packets, its initially uses entropy values in a round-robin fashion until an ACK is received. If the ACK is not ECN-marked, its entropy would be used for the packet that is being sent (Line \#13 in \algo ~\ref{algo:app}). Otherwise, it checks the bitmap in a round-robin fashion. The total paths are dynamically changed with the congestion window (Line \#17 in \algo ~\ref{algo:app}). This is because the path congestion information is fresh within a window. For example, if we maintain a bitmap size of $256$, while the congestion window is one, it requires $256$ RTTs to update the whole bitmap, and the congestion information is already outdated when the round-robin turns to a particular entropy. While checking the bitmap, if the path is ecn-marked, we skip the path and clear the bit for the first skipped path (Line \#23 in \algo ~\ref{algo:app}). 

    \begin{algorithm}[!t]
{\small
    \caption{\sys Adaptive Load Balance}
    \label{algo:app}
    \begin{algorithmic}[1]
    \State \textbf{INIT:} max\_paths=256, bitmap[0...max\_paths]=[0], rr=0, next\_path\_id = -1; 
    \Procedure{update\_ecn\_bitmap}{ecn, path\_id}
       \If{ecn}
   \State next\_path\_id = -(path\_id)
   \State bitmap[path\_id] = 0
   \Else
   \State next\_path\_id = path\_id 
   \State bitmap[path\_id] = 1
   \EndIf
   \EndProcedure
    \Procedure{choose\_path}{}
    \If {next\_path\_id > 0}
    \State rr = next\_path\_id
    \State next\_path\_id = -1
    \Else
    \State flag = false
    \State paths = min(max\_paths, 2*cwnd)
    \State paths = max(8, paths)
    \State rr = (rr + 1)\%paths
    \While{$bitmap[rr]\not=0$}
    \State \Comment{one packet only clears one bit.}
    \If{!flag} 
        \State $bitmap[rr]$ = 0
        \State flag = true
    \EndIf
    \State rr = (rr + 1)\%paths
    \EndWhile
    \EndIf
    \State \textbf{return} $rr$
    \EndProcedure
    \end{algorithmic}
}
\end{algorithm}

\subsection{Congestion Control}
\label{subsec:cc}

\begin{table}[!t]
{\footnotesize
\begin{tabular}{|p{2.3cm}|p{2.3cm}|p{2.3cm}|}
\hline
  \textbf{Constant} & \textbf{Meaning} & \textbf{Recommend
  Values}\\ \hline 
  target\_Qdelay &   & net\_base\_rtt = 12 $\mu s$  \\ \hline
    $target\_Qhigh$ &   a higher target queuing delay threshold & $3$*target\_Qdelay  \\ \hline
    $ewma$ &  Used for the RTT averaging           &$0.125$                     \\ \hline
    bdp\_sf  &  bdp scaling factor          & BDP/(100Gbps*12$\mu s$) \\ \hline
    delay\_sf &   delay scaling factor         & base\_rtt/$12$ $\mu$s  \\ \hline
    $\beta$ &  Used for additive increase          & $5*$MTU*bdp\_sf  \\ \hline
    $\eta$ &   Used for fairness   &  $0.15$*MTU*bdp\_sf   \\ \hline
    $\alpha$&   Used for the RTT gain increase         &  $4.0$*bdp\_sf*
    delay\_sf*MTU/base\_rtt  \\ \hline
    $\gamma$&   Multiplicative decrease parameter         &  $0.8$  \\    
    \hline
\end{tabular}
\caption{Parameters of \sys.}
\label{tbl:ccpara}
}
\end{table}

\begin{algorithm}
{\small
    \caption{\sys Congestion Control}
        \label{algo:cc}
    \begin{algorithmic}[1]
    \Procedure{adjust\_cwnd}{ecn, delay, achievedBDP}    
    \State can\_decrease = now - last\_decrease\_ts $>$ base\_rtt 
    \State can\_fairness\_shuffle = now - last\_selfai\_ts $>$ base\_rtt 
    \State avg\_delay = avg\_delay*(1 - $ewma$) +$ewma$*delay 
    \If{\hbox{!ecn } and delay $>$ $target\_Qhigh$ }
    \State cwnd $\gets$ cwnd + $\frac{\beta}{cwnd}$
    \ElsIf{\hbox{!ecn} and  delay $<$ target\_Qdelay }
    \State cwnd $\gets$ cwnd + $\alpha$*(target\_Qdelay - delay)/cwnd
    \ElsIf{can\_decrease \textbf{and} avg\_delay > target\_Qdelay }
        \If{delay $>$ $target\_Qhigh$ 
        \textbf{and} achievedBDP < max\_cwnd/8}
            \State cwnd = achievedBDP
            \State last\_decrease\_ts  = now
        \ElsIf{delay > target\_Qdelay}
            \State cwnd = cwnd* max( 1 - $\gamma$ * (avg\_delay - target\_Qdelay)/avg\_delay, $0.5$)
            \State last\_decrease\_ts = now
        \EndIf
    \EndIf
    \If{can\_fairness\_shuffle}
    \State cwnd = cwnd + $\eta$
    \State last\_selfai\_ts  = now
    \EndIf
    \State return cwnd
\EndProcedure    
    \end{algorithmic}
    }
        \end{algorithm}    

\begin{algorithm}
{\small
    \caption{\sys AchievedBDP}
    \label{algo:achievedbdp}
    \begin{algorithmic}[1]
    \Procedure{\textbf{update\_achievedBDP}}{ack, now}
    \State can\_clear\_byte = (now -  rxcount\_clear\_ts )  $>$ (base\_rtt + target\_Qdelay)
    \State rx\_count += ack\_for\_probe? 0, acked\_bytes
    \If{can\_clear\_byte}
    \State achievedBDP = rx\_count
    \State rx\_count = 0
    \State rxcount\_clear\_ts = now
    \EndIf
    \EndProcedure
    \end{algorithmic}
}
    \end{algorithm}   
 
\algo~\ref{algo:cc} shows the design of \sys's congestion control algorithm. \tbl \ref{tbl:ccpara} lists the configuration parameters for the performance tuning and use $bdp\_sf$ and $delay\_sf$ as BDP scaling factor and delay scaling factor to adapt to various network speeds and network latency. $target\_Qdelay$ is target queuing delay that we expect the congestion control to be stabilized at. The algorithm maintains one congestion window to handle all the paths congestion. 

For efficiency, \sys increases the rate in two different situations. First, when ecn is not marked and current measured queuing delay (\textbf{delay} in \algo \ref{algo:cc}) is below the target\_Qdelay. This happens when a congestion episode has just passed, the algorithm increases the rate that is proportional to the difference between current RTT and the target\_delay (Line \#8). Secondly, we increase the rate by a constant value when ecn is not marked but measured queuing delay is twice the target delay (Line \#6). This is a situation when this packet incurred a long delay, but the queue length has come down drastically (ecn is not marked), we should increase the rate to avoid link starvation. There are two cases where we reduce the window: depending on the current measured queuing delay is greater than $target\_Qhigh$ or not. If it is greater, we use the achieved BDP (\textbf{achievedBDP} in \algo \ref{algo:achievedbdp}) as our window to quickly converge to a lower rate (Line \#11). Otherwise, we do multiplicative decrease based on how much average measured queuing delay (\textbf{avg\_delay} in \algo \ref{algo:cc}) is over the target\_Qdelay (Line \#14). To ensure fast convergence of fairness, we periodically add a constant small increase to the congestion window (Line \#19). Note that this addition would assist the algorithm to have a slightly bigger window in order for packets to explore different paths.

\subsection{\sys Reliability}
\label{subsec:lr}
Without any advanced features (e.g., trimming or BTS), switches drop packets silently. In \sys, we design a packet loss recovery mechanism that can swiftly detect and retransmit silent packet drops.
In addition, \sys's reliability mechanism also aligns with coalescing ACK mechanism~\cite{smith2010tcp, johnson2018tcp} to accommodate the reality that packet processing rates grow at a slower pace than link speeds.
\sys provides a packet loss detection mechanism across three different time scales: (1) one or two RTTs using out-of-order packets counters; (2) multiple RTTs with probes; (3) tens of RTTs using retransmission timeout. Once a loss is detected, the packet is retransmitted as long as the congestion window allows.   

\subsubsection{Coalescing ACKs and SACKs.}
Upon the arrival of each packet, a receiver sends an Selective Acknowledgment (SACK) packet if one of the following conditions are met: (1) a sufficient number of bytes have been received; (2) a packet with current expected Packet Sequence Number (EPSN) is received; (3) a probing packet is received.
\paragraph{SACK Packet.}
Receiver maintains a bitmap that starts from EPSN, and each bit in the bitmap tracks packet arrivals following EPSN. Packets prior to the expected PSN have been received. 
Due to the constrained bitmap length that can be conveyed in a single SACK packet sent to the sender, the receiver cannot transmit the whole bitmap in one SACK. The complete bitmap is partitioned into segments, each corresponding to the size of a bitmap that a SACK can accommodate. For instance, if the entire bitmap consists of $256$ bits and the SACK bitmap size is $64$ bits, there will be four segments of bitmaps. The receiver should selectively pick the bitmap segment in SACK in order to collectively represent the comprehensive bitmap as efficiently as possible.

As mentioned before, a SACK packet echoes back entropy (i.e., path\_id), ECN mark, and timestamp back to the sender. In addition,
a SACK packet is designed to include the following fields: 
(a) Expected PSN;
(b) Selective ACK Base PSN (Base PSN for Selective ACK bitmap field);
(c) Selective ACK Bitmap;
(d) bytes\_recvd, indicating total received bytes at receive with duplicates eliminated;
(e) Out-Of-Order (OOO) packets counter, representing the number of packets received at the receiver since the EPSN.
These additional fields are shown as in \fig \ref{fig:packet-format}.

The receiver also records the lowest PSN received, LPSN, since the last SACK. The next SACK is sent with the bitmap segment that is associated with LPSN as  retransmission timeout is always triggerred from the lowest unacknowledged PSN. 
Hence we need to update LPSN as early as possible to prevent the sender from timing out. 

\subsubsection{Loss Detection.} 
A sender also maintains a bitmap to keep track of packets that have been cumulatively and selectively acknowledged from the SACK bitmap segments. The sender calculates the inflight bytes as follows:

\begin{align}
    inflight\_bytes &= bytes\_sent - bytes\_recvd \\
    &- bytes\_claimed\_retransmit.
\end{align}

Upon receiving a SACK, the sender updates the inflight bytes based on the SACK information. The inflight should always be smaller than the congestion window. \sys enters loss recovery mode, i.e., loss is inferred, when any of the following conditions meet: (1) the out-of-order packets counter exceeds a threshold; (2) probe-based loss detection; (3) a retransmission timeout occurs. 

\paragraph{OOO-based Detection.} 
Loss is inferred if the out-of-order packets counter carried at SACK exceeds a threshold. The threshold is recommended to be set to the maximum of the current congestion window's worth of packet number and min\_threshold (say 5). This constant threshold prevents spurious retransmissions when the congestion window is less than min\_threshold packets. The rationale for using a threshold based on the congestion window is that if a packet does not get acknowledged within a whole window's time, it is safe to consider it as lost. This also helps us avoid a large bitmap size at receiver/sender 

When the loss detection is triggered, \sys sender enters fast loss retransmission, and records EPSN and current highest received PSN; and considers any unacknowledged packets between EPSN and the highest received PSN are lost.

As long as the congestion window allows, the sender retransmits unacknowledged packets starting from EPSN. Upon receiving new SACKs, it updates EPSN and SACK bitmap and sends the remaining unacknowledged packets. The process continues until all the packets between the original EPSN and the recorded highest received PSN are ACKed. Then, the sender exits the loss recovery mode. 

\paragraph{Probe-based Detection.}
A probe packet is sent when the sender does not get any SACK over a specific duration, ideally set as $n$ (e.g. n = 3) network wide base RTTs. A probe packet can indicate the SACK bitmap to be returned using a specific SACK base. In common cases, SACK base will be EPSN. Upon receiving a probe packet, the receiver is expected to issue a SACK using the SACK base immediately. A dedicated bit is used to indicate that the SACK is triggered by the probe packet.

If an SACK packet for probe packet comes back and the measured RTT is small, and no other ACK is received after the probe is sent (Line \#10 in \algo \ref{algo:overview}), the sender gets into the loss recovery mode and using the same way as ooo-based method to recover from packet losses.   
In case the probe packets are also lost, the sender will send another probe packet transmission if it doesn't receive the SACK within the specified period, $n$ times network wide base RTTs. 

\paragraph{Timeout-based Detection.}  
A single timestamp per flow is maintained at the sender to monitor retransmission timeouts. This timestamp is initialized upon the creation of the connection and is updated whenever a higher EPSN is received. If the timeout period has passed without the sender receiving the acknowledgement for current EPSN packet, all the unackowledged packets are declared lost and retransmitted.  The timeout value is generally configured in hundreds of microseconds or, at the very least, tens of microseconds.

%% file: sections/5evaluation.tex
\begin{figure*}[!htb]%
\includegraphics[width=0.21\linewidth]{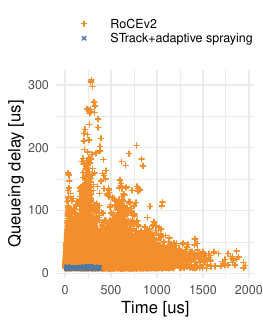}%
\includegraphics{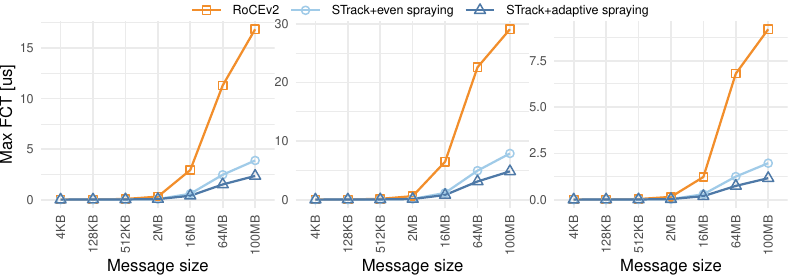}\hfill
    \begin{minipage}{\textwidth}
    \vspace{-1em}
        \begin{minipage}{0.24\textwidth}
            \caption{\small Switch link queuing delay.}
            \label{fig:isl-queuing-delay}
        \end{minipage}\hfill
        \begin{minipage}{0.24\textwidth}
            \caption{\small8K nodes permutation, 400Gbps.}
            \label{fig:perm-400}
        \end{minipage}\hfill
        \begin{minipage}{0.24\textwidth}
            \caption{\small8K nodes permutation, 200Gbps.}%
            \label{fig:permu-200}%
        \end{minipage}\hfill
        \begin{minipage}{0.24\textwidth}
            \caption{\small8K nodes permutation, 800Gbps.}%
            \label{fig:permu-800}%
        \end{minipage}\hfill
    \end{minipage}
\end{figure*}%

\begin{figure*}[!htb]
    \includegraphics{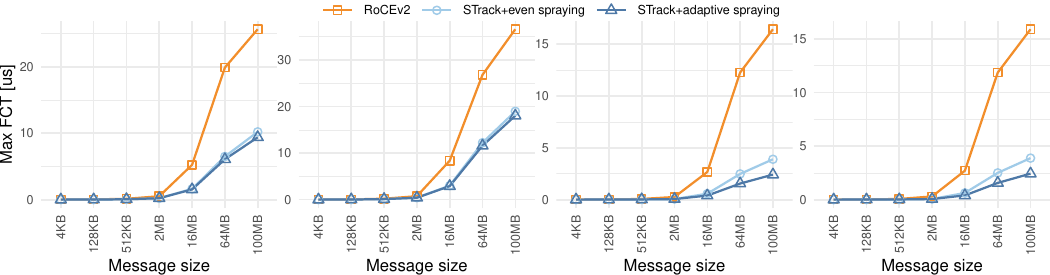}
    \begin{minipage}{\textwidth}
        \vspace{-1em}
        \begin{minipage}{0.24\textwidth}
            \caption{\small 8K nodes permutation, 4:1 over-subscription.}
            \label{fig:4oversub}
        \end{minipage}\hfill
        \begin{minipage}{0.24\textwidth}
            \caption{\small 8K nodes permutation, 8:1 over-subscription.}
            \label{fig:8oversub}
        \end{minipage}\hfill
        \begin{minipage}{0.24\textwidth}
            \caption{\small 8K nodes permutation, 64 links down.}
            \label{fig:64linkdown}
        \end{minipage}\hfill
        \begin{minipage}{0.24\textwidth}
            \caption{\small 8K nodes permutation, 256 links down.}
            \label{fig:256linkdown}
        \end{minipage}%
    \end{minipage}
\end{figure*}

\section{Evaluation}
We evaluate \sys through large-scale simulations using synthetic traffic and collective workloads to answer the following questions:
\begin{itemize}

 \item How does \sys's \adptivespray impacts the tail latency (\subsec \ref{subsec:perm})?
 \item How does \sys's joint optimization of adaptive load balancing and congestion control perform under over-subscribed as well as asymmetric networks (\subsec \ref{subsec:oversub})?
 \item How does \sys handle loss recovery upon high incasts and what is its impact on switch buffer occupancy (\subsec \ref{subsec:incast})? 
 \item How does \sys impact the AI/ML workloads' communication performance (\subsec \ref{subsec:ar})? 
\end{itemize}
\subsection{Experiment Setup}
We compare \sys with RoCEv2~\cite{rocev2}, which is widely deployed transport over Ethernet in AI/ML network.
We configure the constants for \sys congestion control as suggested in ~\tbl ~\ref{tbl:ccpara} and set the maximum number of paths as $256$ for adaptive load balancing and switch ECN mark threshold $K_{min} = 25\% BDP$ and $K_{max} = 75\% BDP$. Switch is operated without any advanced features enabled, and drops packets if queuing exceeds $5$ BDP. 
\roce utilizes DCQCN~\cite{dcqcn} as its primary congestion control mechanism and supports single-path operation. Additionally, it necessitates Priority Flow Control (PFC) to be enabled throughout the entire network to prevent packet loss caused by buffer overflow. 
We implemented shared buffer architecture for PFC to best demonstrate the performance of \roce. We configured the total switch buffer size to be $256$MB for a total capacity of $51.2$Tbps and scaled the buffer size accordingly for switches with different radices. For instance, the switch buffer size is set to 128MB for a 25.6Tbps switch.
We set the ECN
threshold to one BDP for DCQCN and configure Lossless for RoCEv2. 

Switch has ECMP~\cite{ecmp} as the routing strategy. We extend htsim~\cite{htsim} simulator to implement the RoCEv2 and \sys as well the switch models. 

\paragraph{Performance metrics.}
For the synthetic workloads, we use tail flow completion time (max FCT) as our application performance metric. The FCT is measured from the time the message is ready to be pushed to the networking stack to the time when the last packet of that message gets acknowledged at the sender. 
For the collective workloads, we measure the collective completion time, which is measured when the first message of the collective sent to the time when the last message finishes.

\subsection{Synthetic Traffic}
We use a 2-tier standard fattree topology~\cite{eflops, accl, hpn} with $8192$ nodes for the evaluation with MTU size of $4$KB. Our tests vary the link speed for $200$G, $400$G and $800$G and the default link speed is $400$G. The synthetic workloads include permutation traffic pattern and incast traffic pattern.
The network topology is configured with (1) full-bisectional network; (2) over-subscription ratio of $4$:$1$ or $8$:$1$; (3) asymmetric network. The asymmetric network is created by disabling 0.78\% or 3.1\% of inter-switch links. The network wide RTT is $8\mu$s, and thus, the network bandwidth-delay-product (BDP) for $400$Gbps is $400$KB. The message size varies from $4KB$, $128KB$, $512KB$ $2MB$, $16MB$, $64MB$ and $100MB$.

\begin{figure*}[!htb]
        \begin{minipage}{0.32\textwidth}
        \centering
        \includegraphics{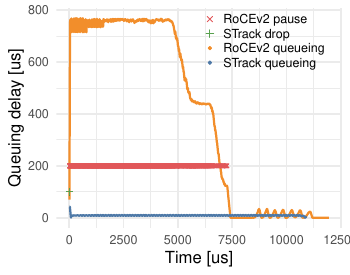}
 		\caption{\small 32->1 incast, Queuing delay, drop and pause as time goes.}
           \label{fig:32-1incastqueuing}
    \end{minipage}\hfill 
    \begin{minipage}{0.32\textwidth}
        \centering
        \includegraphics[trim=0mm 0mm 168pt 0mm, clip]{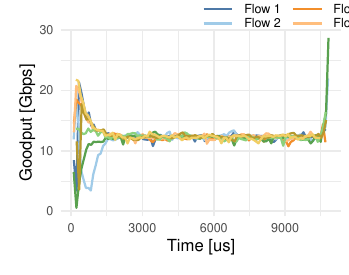}
 		\caption{\small 32->1 incast, \sys throughput over time.}
        \label{fig:strack-tput}
    \end{minipage}\hfill
    \begin{minipage}{0.32\textwidth}
        \centering
        \includegraphics[trim=167pt 0mm 0pt 0mm, clip]{evalfigs/figure10-11.pdf}
 		\caption{\small 32->1 incast, \roce throughput over time.}
        \label{fig:roce-tput}
    \end{minipage}\hfill
        \begin{minipage}{0.5\textwidth}
        \centering
        \includegraphics[width=\linewidth]{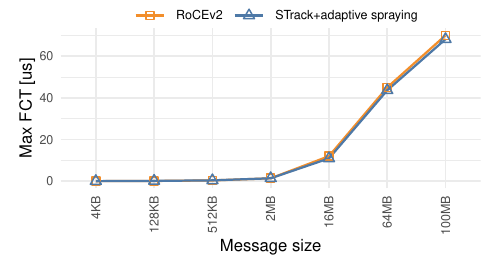}
 		\caption{\small 32->1 incast, message completion time}
        \label{fig:32-1incast}
    \end{minipage}\hfill
            \begin{minipage}{0.5\textwidth}
        \centering
        \includegraphics[width=\linewidth]{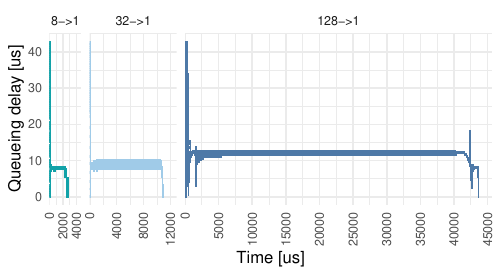}
     	\caption{\small \sys last-hop switch queuing delay for 8->1 incast, 32->1 incast, 128->1 incast as time goes.}
        \label{fig:queuing-incast-8-16-128}
    \end{minipage}\hfill
\end{figure*}

\paragraph{Permutation.}
\label{subsec:perm}

We utilize the permutation traffic pattern to demonstrate effective load balancing. This pattern is created by randomly pairing two nodes for each flow, ensuring that any given node serves as the receiver for one flow and the sender for another. For this traffic pattern, if the network load balancing is perfect or nearly perfect, congestion control should not reduce the window size. This workload is ideal for showcasing the effectiveness of the network load balancing scheme, and demonstrate how congestion control interacts with load balancing. Given there are a total of $8$K nodes in the network topology, the permutation workload has a total of $8K$ flows and we report the maximum flow completion time over the $8K$ flows.
Figure \ref{fig:perm-400} shows the maximum flow completion time of the various message size for \sys, \sys's congestion control with \evenspray and \roce.

Figure \ref{fig:perm-400} shows the max flow completion time of the various message size for \sys, \sys's congestion control with \evenspray and \roce with the default link speed of $400$G.
\sys improves the max FCT over \roce up to $6.3$X, as \roce traffic uses one path and is highly impacted by hash collisions. As a result, \roce is not able to leverage the idle parallel paths to achieve higher throughput. In addition, adaptive packet spray improves the max FCT over \evenspray up to 33\%, when the message size is $100$MB.
\sys's \adptivespray requires at least one RTT to get the congestion signal, i.e., ECN, back to sender to skip the bad path. Thus, the performance differentiation for \evenspray and \adptivespray starts to show only for larger message sizes, e,g, after 16MB. 

Figure \ref{fig:isl-queuing-delay} samples the switch queuing delay as the simulation time progresses in the $16MB$ experiment in \fig \ref{fig:perm-400} for \sys and \roce. There are a total of $16K$ switch queues, $8$K queues from ToR to Spine Switch and the other $8$K queues from Spine to ToR. To avoid too many logs to be processed, we only generate logs if the queuing delay is larger than $8$us (the base RTT). Hence, if we do not see any points at one time moment in \fig \ref{fig:isl-queuing-delay}, it means that across the $16$K queues, there is not any queue whose queue depth exceeds the BDP at that time point. \fig \ref{fig:isl-queuing-delay} shows that, initially before $370$us, \sys's \adptivespray experiences mild congestion due to hash collisions, but it adapts to skip the bad paths, and finds the good paths. Hence after $370$us, all the flows settle down to the good paths, and thus, 
 the switch queuing delays do not exceed $8$us after $370$us.
 In contrast, \roce continues experiencing hash collisions, which leads to the queuing delay up to $250$us until the end of experiment.

To demonstrate the robustness of \sys, we also run the same workloads with link speeds of $200$Gbps and $800$Gbps. \fig \ref{fig:permu-200} and \ref{fig:permu-800} shows the maximum FCT for different message sizes for $200$Gbps and $800$Gbps, respectively. \sys \adptivespray outperforms \evenspray as message size increases. \sys \adptivespray consistently outperforms \roce starting from message size $128$KB and gets the performance improvement by up to $6$X when the message size is $64$MB and $100$MB.

\paragraph{Permutation with over-subscribed network.}
\label{subsec:oversub}
Network over-subscription is common in data center networks, as it can reduce the cost for the network devices, e.g., switches and links. We change the full bi-sectional 2-tier network topology to be the $4$:$1$ or $8$:$1$ oversubscribed, where the total bandwidth from the ToR switches to the spine switches is $1/4$ or $1/8$ of total bandwidth from host to ToR switches. We use the permutation traffic pattern as the workloads. This setup evaluates the effectiveness of joint optimization of congestion control and multipath load balancing mechanism. 

Figures \ref{fig:4oversub} and \ref{fig:8oversub} display the maximum flow completion time (FCT) for various message sizes using \sys and \roce in 4:1 and 8:1 over-subscription networks, respectively. \sys consistently outperforms \roce by up to $3$X across all message sizes. \sys's \adptivespray mechanism effectively balances the load across inter-switch links, fully utilizing all available path bandwidth. Additionally, its congestion control reduces the window size by factors of $4$ or $8$ to accommodate the 4:1 or 8:1 over-subscription. In contrast, \roce's single path strategy suffers from heavy hash collisions, leading to network congestion, a reduction in sending rate by DCQCN, and under utilization of the network. 
As congestion control plays the main role in driving the performance in over-subscription scenarios, we observe that \adptivespray and \evenspray exhibit similar behavior due to their use of the same congestion control algorithm.

\paragraph{Permutation with Asymmetric network.}
\label{subsec:failedlinks}

Network link failures are common in data centers~\cite{hpn}. To simulate an asymmetric network, we disconnect the links between ToR and spine switches. We fixed the number of switches experiencing link failures at $16$ and disconnected either 64 (0.78\%) or 256 (3.1\%) out of a total of 8,192 inter-switch links. This corresponds to $4$ links per ToR switch for the $64$-link down scenario and $16$ links per ToR switch for the 256-link down scenario. We ran the same permutation traffic pattern on this network to validate the efficiency of \sys multipath load balancing.

Figures \ref{fig:64linkdown} and \ref{fig:256linkdown} show the maximum flow completion time for an $8K$-flow permutation with 64 and 256 link failures. Both \evenspray and \adptivespray consistently outperform \roce's single-path transport by up to $6$X, particularly when 64 links are down and the message size is $100$MB. We do not observe a significant performance decrease despite these link failures. Additionally, \adptivespray gains up to a 60\% performance improvement over \evenspray when $64$ links are down and the message size is $100$MB.

\paragraph{Incast.}
\label{subsec:incast}

We run experiments for $7$ message sizes the same as \fig \ref{fig:perm-400}-\ref{fig:256linkdown} for the 8-to-1 incast, 32-to-1 incast, and 128-to-1 incast, respectively. 
\fig~\ref{fig:32-1incastqueuing} shows the last-hop switch queuing delay, packets drop for \sys and pause generation time for \roce as time goes in a $32$-to-$1$ incast experiment with $16$MB. 
\roce shows persistently high switch queuing delay up to $7500$us (worth of $37.5$MB switch queue depth) for $5000$us, and takes $7800$us to converge to the right rate, while \sys takes $95$us to reach to the stable state, which shows the fast convergence of \sys. We also label the pause generation time (marked red) and packets drop time (marked green) in \fig~\ref{fig:32-1incastqueuing}.
The switch keeps sending the pause frames until the convergence point with \roce,  which is due to the slow convergence of DCQCN. Because of its fast convergence, \sys only drops packets at the first RTT, and there is not any more packet drops after the first RTT.

\fig \ref{fig:strack-tput} and \fig \ref{fig:roce-tput} show the instantaneous throughput of $8$ flows, averaged over $100$ microsecond intervals, for \sys and \roce respectively, same as the experiment depicted in Figure \ref{fig:32-1incastqueuing}. Figure \ref{fig:strack-tput} demonstrates that the $8$ flows with \sys can achieve their fair share of a bottleneck link, and converge to a new, stable bandwidth share. Conversely, \roce exhibits fairness issue in \fig \ref{fig:roce-tput}, with PFC pauses, whose on and off behavior also prevents each flow from quickly attaining its fair share.

\fig \ref{fig:32-1incast} shows the maximum flow completion time of \sys and \roce for each message size with 32->1 incast.  As
\roce relies on lossless link support, the bottleneck bandwidth of the last hop is fully utilized. Although pauses are generated, they tend to cause collateral damage to victim flows that share the uplink bandwidth. For the flows that go to the incast destination, \roce can achieve near-optimal flow completion time. As a lossy protocol, \sys incurs losses during an incast event, especially for the first RTT. The challenge is to recover quickly from these losses. As shown in \fig \ref{fig:32-1incast}, \sys can match \roce's performance by finishing about the same time across various message sizes, indicating that \sys's fast packet loss recovery mechanism allows its end-to-end performance to align with that of the lossless network.

\fig \ref{fig:queuing-incast-8-16-128} shows the  queuing delay of last-hop ToR switch for $8$->$1$, $32$->$1$, $128$->$1$ incast experiments with message size of $16$MB. \sys is able to stabilize at the target queuing delay across different incast degrees, which shows the robustness of the \sys's congestion control. Note that the stabilized point of 128->1 incast is is slightly higher than the target queuing delay, i.e., $8\mu$s, this is because the ideal congestion window in this case is $1.3$ packets. Due to rounding error, the inflight packets tend to vary between 1 or 2 packets, which causes slightly elevated queueing delay.

\begin{figure*}[!htb]
    \includegraphics{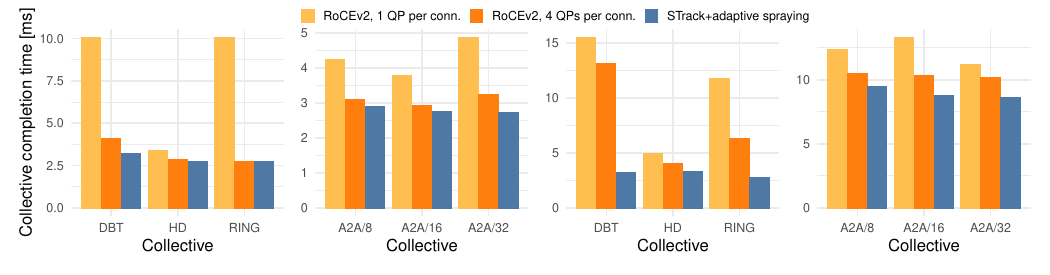}
    \begin{minipage}{\textwidth}
        \vspace{-1em}
        \begin{minipage}{0.24\textwidth}
            \caption{\small 64 AllReduce Collectives, full bisectional network.}
            \label{fig:ar-64jobs}
        \end{minipage}\hfill
        \begin{minipage}{0.23\textwidth}
            \caption{\small 64 AlltoAll Collectives, full bisectional network.}
            \label{fig:a2a-64jobs}
        \end{minipage}\hfill
        \begin{minipage}{0.24\textwidth}
            \caption{\small 64 Allreduce Collectives, 4:1 oversubscribed network.}
            \label{fig:ar-64jobs-oversub4}
        \end{minipage}\hfill
        \begin{minipage}{0.24\textwidth}
            \caption{\small 64 AlltoAll Collectives, 4:1 oversubscribed network.}
            \label{fig:a2a-64jobs-oversub4}
        \end{minipage}\hfill
    \end{minipage}
\end{figure*}

\begin{figure*}[!htb]
    \includegraphics{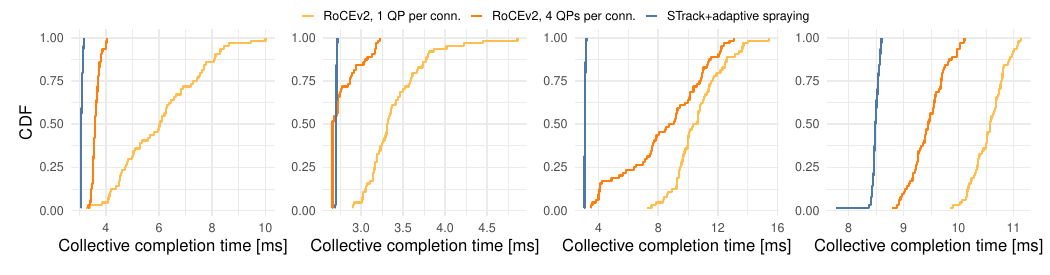}
    \begin{minipage}{\textwidth}
        \vspace{-1em}
        \begin{minipage}{0.23\textwidth}
            \caption{\small DBT, full bisectional network.}
            \label{fig:cdf-dbt-64jobs}
        \end{minipage}\hfill
        \begin{minipage}{0.23\textwidth}
            \caption{\small 32, A2A, full bisectional network.}
            \label{fig:cdf-a2a-32-64jobs}
        \end{minipage}\hfill
        \begin{minipage}{0.23\textwidth}
            \caption{\small DBT, 4:1 oversubscribed network.}
            \label{fig:cdf-dbt-64jobs-oversub4}
        \end{minipage}\hfill
        \begin{minipage}{0.23\textwidth}
            \caption{\small 32, A2A, 4:1 oversubscribed network.}
            \label{fig:cdf-a2a-32-64jobs-oversub4}
        \end{minipage}\hfill
    \end{minipage}
\end{figure*}

\subsection{AI/ML Workloads}
\label{subsec:ar}
Collective algorithms in machine learning training systems are designed to avoid network incast [11, 12, 27]. However, network imbalance and system jitter can affect the ideal message scheduling and degrade system performance. In this section we evaluate \sys on a representative set of ML workloads. To model the communication scheduling of Deep
Learning (DL) training workloads, especially  collectives~\cite{nccl, rccl}, we generate a message trace based on the three AllReduece algorithms, i.e., DoubleBinaryTree (\textbf{DBT}), Ring (\textbf{RING}), and HalvingDoulbing (\textbf{HD}), and a AlltoAll collective (\textbf{A2A}).

The AllReduce traces express message dependency, where messages from later steps are sent only after messages in previous steps are received. For instance, in the \dbt algorithm, a node in the middle layer of the tree only sends messages to its parent after receiving messages from its two children. The AllToAll traces are sequenced to prevent incast. For example, the first message from rank $n$ is sent to rank $(n+1)\%N$, and the second message at rank $n$ is sent to rank $(n+2)$\%$N$.
For AlltoAll communication, we restrict the number of active parallel connections at both the sender and receiver to various levels, such as $16$, $32$, and $64$. This helps to control the degree of incast and outcast from a network perspective.

The simulated topology is a full-bisection 2-tier Clos network with all links operating at the same speed as the default. We vary the link speed to demonstrate the robustness of \sys. We only show the results for the network link speed of $400$Gbps as the experiment results of different link speeds are similar. We assume one NIC maps to one GPU. 
Each message chunk size is $128$KB to utilize the pipeline. 
We primarily assess a multi-job scenario involving 64 identical collectives running concurrently on a 2048-NIC cluster, with each collective group randomly distributed throughout the cluster.

\paragraph{Full Bisectional Network.}
\fig \ref{fig:ar-64jobs} displays the maximum collective finishing time for $64$ allreduce collectives using different allreduce algorithms with \sys and \roce on a full bisectional network. \sys with \adptivespray outperforms \roce with 1 \qpsperconn by up to $3$x and \roce with 4 \qpsperconn by up to $27.4\%$, especially with the DoubleBinaryTree algorithm. The DoubleBinaryTree algorithm exhibits a 2:1 incast property, where the limitations of DCQCN become evident. \roce with 4 \qpsperconn performs better than \roce with 1 \qpsperconn across various allreduce algorithms, as the increased entropy improves performance, albeit with higher CPU cost. \fig \ref{fig:cdf-dbt-64jobs} plots the CDF of $64$ collective finishing time of DoubleBinaryTree of \fig \ref{fig:ar-64jobs} experiment. It shows that the $64$ collectives achieves similar finishing times with \sys, while there is $26.6\%$ difference between the smallest finishing time and the maximum finishing time with \roce with 4 \qpsperconn, which indicates \sys not only has better tail latency of each collective, but also better fairness compares with \roce. 
\fig \ref{fig:a2a-64jobs} shows the maximum collective finishing time for $64$ AlltoAll collectives across different levels of parallelism using \sys and \roce. \sys with \adptivespray outperforms \roce with $1$ \qpsperconn by up to $79.2\%$ and \roce with $4$ \qpsperconn by up to $18.9\%$, particularly at a parallel degree of $32$. \fig \ref{fig:cdf-a2a-32-64jobs} shows the CDF of finishing times for 64 alltoall collectives with a parallel degree of 32. This further demonstrates that \sys offers better fairness compared to \roce.

\paragraph{Oversubscribed Network.}
\fig \ref{fig:ar-64jobs-oversub4} shows the maximum collective finishing time for $64$ allreduce collectives using different allreduce algorithms with \sys and \roce on a $4$:$1$ oversubscribed network. \sys with \adptivespray outperforms \roce with 1 \qpsperconn by up to $4.86$X and \roce with 4 \qpsperconn by up to $4.13X$, especially with the \dbt algorithm. 
We observed that the collective completion time of \sys does not significantly increase in the oversubscribed network compared to that in the full bisectional network (Figure \ref{fig:ar-64jobs}). This indicates that the traffic generated by the allreduce algorithm does not fully utilize the full bisectional network.
However, in this scenario, the RING and HalvingDoubling configurations of \roce with 1 \qpsperconn and 4 \qpsperconn exhibit worse performance than that of the full bisectional bandwidth due to its poor load balancing and congestion control mechanisms. Although \roce with 4 \qpsperconn performs slightly better than \roce with 1 \qpsperconn across various allreduce algorithms, it does not match \sys's load balancing efficiency, triggering DCQCN to cut rate that leads to poor performance. 
\fig \ref{fig:cdf-dbt-64jobs-oversub4} shows the CDF of finishing times for $64$ allreduce collectives with \dbt algorithm. Comparing to \roce, \sys shows much lower collective finishing time for each collective and offers better fairness.

\fig \ref{fig:a2a-64jobs-oversub4} shows the maximum collective finishing time for $64$ AlltoAll collectives across different levels of parallelism using \sys and \roce on a $4$:$1$ oversubscribed network. \sys with \adptivespray outperforms \roce with $1$ \qpsperconn by up to $51.5\%$ and \roce with $4$ \qpsperconn by up to $17.8\%$, particularly at a parallel degree of $16$. Compared to the allreduce collectives (\fig \ref{fig:ar-64jobs}), alltoall collectives require higher bandwidth for inter-switch links. Consequently, we observe approximately four times the maximum collective finishing time in the full bisectional network scenario. \fig \ref{fig:cdf-a2a-32-64jobs-oversub4} displays the CDF of finishing times for 64 alltoall collectives with a parallel degree of $32$ for \sys and \roce. 
It demonstrates that \sys completes one all-to-all collective in $7.78$ ms, while other collectives complete in 8.5 ms. This is because $16$ of the $32$ NICs for the earlier-finishing collective are located on the same ToR, thereby reducing its traffic to the bottleneck inter-switch links by one-fourth. As a result, \sys consistently shows superior performance and greater fairness compared to \roce.

%% file: sections/6related.tex
\section{Related Work}


Our work draws on the long history of congestion control and multipath load balancing mechanisms. Below, we classify various congestion control techniques based on the type of congestion signal they employ and their approaches to regulating transmission rates, distributing network traffic across multiple paths, and achieving joint optimization of these aspects.

\subsection{Congestion Signals}

The choice of the signal to use to detect congestion largely impacts the \textit{\textbf{responsiveness}} of the algorithm.

\paragraph{Delay-based.}

The simplest signal is measured RTT, where either the sender measures the round trip time (e.g., Swift~\cite{swift}, Timely~\cite{timely}, TCP Vegas~\cite{vegas}) or the receiver annotates ACK packets with one-way delay information (e.g., OnRamp~\cite{onramp}). Accurate RTT measurement is critical and may require special OS or hardware support such as Snap in Swift~\cite{swift}. Delay-based signals are often described as \textit{``multi-bit''} as they provide rich information for exact queuing delay along the path in the network.

\paragraph{Explicit Congestion Notification (ECN).}

ECN is widely used in datacenters and is available in all modern switches.
Algorithms like DCTCP~\cite{dctcp}, Hull~\cite{hull}, and D$^2$TCP~\cite{D2TCP} all rely on ECN marking. For example, DCQCN~\cite{dcqcn} combines ECN and \textit{Priority Flow Control} (PFC) to react to congestion. However, although egress-marked ECN-based algorithm can react quicker than delay-based ones~\cite{ecn_or_delay}, they perform poorly when dealing with incasts and are usually hard to tune~\cite{hpcc}. 

\paragraph{In-Network Telemetry.} While ECN is a simple and widely available example for in-network telemetry (INT), other protocols rely on more complex signaling such as maximum queue occupancy along the path~\cite{wangposeidon}. Modern switches offer a plethora of such signals and protocols using those are in active development. HPCC~\cite{hpcc}, for example uses precise load information along the path to adjust its rate at the sender.

\paragraph{Packet-Trimming and Back-to-Sender.}

Contemporary programmable switches offer queuing stats at their ingress points, enabling the application of cutting payload techniques to data packets~\cite{cp, ndp}. Additionally, switches can duplicate header packets and send them back to the endpoints~\cite{sfc, bolt}. This approach requires the notification of network congestion status to endpoints even without experiencing ongoing congestion.



\subsection{Congestion Management}

\subsubsection{Congestion Control Algorithm.}

The decision on how to react to congestion can occur in different parts of the network, and this has a direct impact on the \textbf{\textit{visibility}} and on which type of congestion the algorithm can react to.

\paragraph{Receiver-based.} Some algorithms exploit the knowledge of the receiver about concurrently incoming flows to schedule the transmission by the different senders directly. Many of these schemes such as NDP~\cite{ndp}, EQDS~\cite{eqds}, pHost~\cite{phost}, ExpressPass~\cite{expresspass}, and Homa~\cite{homa} use end-to-end credits back to the sender to control flows rates. However, whereas these schemes can deal very well with incast traffic, they usually fall short when dealing with congestion happening in the network fabric~\cite{ndp, eqds}.

\paragraph{Sender-based.}
Algorithms such as Swift~\cite{swift}, DCQCN~\cite{dcqcn}, DCTCP~\cite{dctcp}, HPCC~\cite{hpcc}, Timely~\cite{timely}, Poseidon~\cite{wangposeidon}, and others instead  rely on the sender to adjust its transmission rate based on congestion information received from the receiver.
In some schemes like Bolt~\cite{bolt}, switches can directly indicate congestion back to the sender and thus shortcut the path to the receiver.
These sender-based congestion control algorithms can respond to congestion occurring anywhere in the network, which is why they are widely used in data center networks today.

\subsubsection{Multipath load balancing.}

The multipath load balancing mechanisms can be categorized into host-driven solutions~\cite{hermes, mprdma, presto}, switch-driven solutions~\cite{conga, detail, drill} and centralized solutions~\cite{hedera, mahout, microTE, FastPass}. Here, we focus on closely host-driven related works.

Today's data centers primarily rely on single-path transport methods, such as TCP~\cite{tcp} and RoCEv2~\cite{rocev2}, which are standard practices. However, these methods often encounter well-documented ECMP hash collisions. In contrast, MPTCP~\cite{mptcp} addresses these challenges by breaking down a single flow into multiple sub-flows and intelligently distributing packets among these sub-flows based on network congestion conditions.
Presto~\cite{presto} breaks a flow even further into flowcells and picks paths in a round-robin fashion.

Another approach, known as Hermes~\cite{hermes}, leverages ECN and RTT to differentiate among good, gray, and bad paths, although it depends on pre-installed switch routing to enforce specific paths. Lastly, Protective Load Balancing (PLB~\cite{plb}) utilizes RTT and ECN to detect ECMP hash collisions and dynamically adjusts the flow entropy field to avoid congested paths.

\subsubsection{Joint Optimization.}

There have been limited prior work focused on jointly optimizing congestion control and multipath load balancing, which is the primary objective of our system. One example of such an approach is MP-RDMA~\cite{mprdma}, which employs packet spraying during the initial window, and handles fabric and incast congestion using ECN only. As a result, the algorithm converges slowly as it requires multiple RTTs to reach stability upon a congestion event.
In a similar vein, Scalable Reliable Datagram (SRD~\cite{srd}) utilizes congestion control techniques similar to BBR~\cite{bbr} and incorporates multipath load balancing through packet spraying. However, as a proprietary scheme, it is not clear the detailed algorithms and its loss recovery scheme.

\sys utilizes ECN for path selection and switches to paths that are devoid of ECN markings. It combines ECN and RTT in the congestion control algorithm to effectively tackle network congestion. 

%% file: sections/7con.tex
\section{Conclusion}
We tackle the challenge of managing collective communication for AI/ML workload without advanced switch supports, such as packet trimming or in-network-telemetry. We propose and demonstrate that \sys, a NIC-side only and hardware-offloaded reliable transport protocol, can achieve ultra-low latency, very high bandwidth utilization and well balanced networks. By taking advantage of egress-marked ECN, 
\sys swiftly changing path upon early congestion detection and adaptively load balancing to utilize the overall available capacity in a multipath network. Congestion window control only comes in when a majority of the paths are congested when average RTT is elevated.
With a unique error recovery mechanism, \sys also ensures rapid packet recovery within a multipath context. Our evaluation results show that highly utilized and well balanced network links are the key to ideal performance for AI/ML workloads.
\sys outperforms \roce up to $6$X with synthetic workloads and by $27.4$\% with collective workloads, even with the optimized \roce system setup.

%% file: sections/acknowledgement.tex
\section*{Acknowledgments}
We would like to thank Weijia Yuan, Gaurav Borker 
for their thoughtful feedback.